\RequirePackage{lineno}
\documentclass[aps,twocolumn,superscriptaddress,preprintnumbers,amsmath,amssymb]{revtex4}
\usepackage{graphicx,color,dcolumn,booktabs,bm}
\usepackage{longtable,lscape}
\usepackage{amsmath}
\usepackage{indentfirst}
\usepackage{epsfig}
\usepackage{feynmf}
\usepackage{slashed}
\usepackage{cases}
\usepackage{color}
\usepackage{multirow}
\usepackage{graphicx,color,dcolumn,booktabs,bm}
\usepackage{verbatim}
\usepackage[colorlinks,linkcolor=blue,anchorcolor=green,citecolor=blue]{hyperref}
\usepackage{mathrsfs}
\usepackage{hyperref}
\usepackage{rotating}
\usepackage{threeparttable}
\usepackage{anyfontsize}
\usepackage{xcolor}
\usepackage{float}
\usepackage{epsfig}
\usepackage{rotating}
\usepackage{ulem}
\usepackage{amssymb}
\usepackage{amsmath}
\usepackage{txfonts}
\usepackage{upgreek}
\usepackage{mathtools}
\usepackage{tabularx}
\usepackage{booktabs}
\usepackage{titlesec}
\titlespacing*{\section}
  {0pt}
  {1\baselineskip}
  {1\baselineskip}

\titleformat{\section}
  {\normalfont\bfseries}
  {\arabic{section}.\ }
  {0em}
  {}

\usepackage{soul} 
\usepackage{lineno}

\usepackage[capitalize]{cleveref}
\definecolor{nicered}{rgb}{.7,.1,.1}
\definecolor{nicegreen}{rgb}{.1,.5,.1}
\definecolor{darkblue}{rgb}{0,0,.5}
\hypersetup{colorlinks, citecolor=nicegreen,linkcolor=nicered, urlcolor=darkblue}
\usepackage{multirow}
\usepackage{slashed}
\usepackage{verbatim}

\providecommand{\Keywords}[1]{\textbf{Keywords}:#1}

\usepackage{xspace}

\begin{document}

\title{Novel method for identifying the heaviest \mbox{QED} atom}

\author{Jing-Hang~Fu}
\thanks{These authors contributed equally to this work.}
\affiliation{School of Physics, Beihang University, Beijing 100083, China}

\author{Sen~Jia}
\thanks{These authors contributed equally to this work.}
\affiliation{School of Physics, Southeast University, Nanjing 211189, China}

\author{Xing-Yu~Zhou}
\affiliation{School of Physics and Electronic Technology, Liaoning Normal University, Dalian 116029, China}

\author{Yu-Jie~Zhang}
\thanks{zyj@buaa.edu.cn (Corresponding author)}
\affiliation{School of Physics, Beihang University, Beijing 100083, China}
\affiliation{Peng Huanwu Collaborative Center for Research and Education, Beihang University, Beijing 100191, China}

\author{Cheng-Ping~Shen}
\thanks{shencp@fudan.edu.cn (Corresponding author)}
\affiliation{Key Laboratory of Nuclear Physics and Ion-beam Application (MOE) and Institute of Modern Physics, Fudan University, Shanghai 200443, China}
\affiliation{School of Physics, Henan Normal University, Xinxiang 453007, China}

\author{Chang-Zheng~Yuan}
\thanks{yuancz@ihep.ac.cn (Corresponding author)}
\affiliation{Institute of High Energy Physics, Chinese Academy of Sciences, Beijing 100049, China}
\affiliation{University of Chinese Academy of Sciences, Beijing 100049, China}

\begin{abstract}
\mbox{QED} atoms are composed of unstructured and point-like lepton pairs bound together by the electromagnetic force. The smallest and heaviest \mbox{QED} atom is formed by a $\tau^+\tau^-$ pair. Currently, the only known atoms of this type are the $e^+e^-$ and $\mu^+e^-$ atoms, which were discovered 64 years ago and remain the sole examples found thus far.
We demonstrate that the $J_\tau$ ($\tau^+\tau^-$ atom with $J^{PC}=1^{--}$) atom signal can be observed with a significance larger than $5\sigma$ including both statistical and systematic uncertainties, via the process $e^+e^-\to  X^+ Y^- \slashed{E}$ ($X,\,Y=e$, $\mu$, $\pi$, $K$, or $\rho$, and $\slashed{E}$ is the missing energy due to unobserved neutrinos) with $1.5~{\rm ab^{-1}}$ data taken around the $\tau$ pair production threshold. The $\tau$ lepton mass can be measured with a precision of 1~keV with the same data sample. This is within one year's running time of the proposed super tau-charm facility in China or super charm-tau factory in Russia.

\vspace{0.3cm}
\noindent
\Keywords{~\mbox{QED} atom, Tau lepton, $e^+e^-$ annihilation, STCF, SCTF}
\end{abstract}

\maketitle

\nopagebreak

\section{Introduction}
Quantum electrodynamics (\mbox{QED}) atoms are formed of lepton pairs ($e^+ e^-$, $\mu^+e^-$, $\tau^+e^-$, $\mu^+\mu^-$, $\tau^+\mu^-$, and $\tau^+\tau^-$) by electromagnetic interactions, similar to hydrogen formed of a proton and an electron. Their properties have been studied to test \mbox{QED} theory~\cite{Karshenboim:2005iy,Safronova:2017xyt}, fundamental symmetries~\cite{Hughes:2001yk,Yamazaki:2009hp}, gravity~\cite{MAGE:2018wxk,Soter:2021xuf}, and search for physics beyond the Standard Model~\cite{Bernreuther:1988tt,Yamazaki:2009hp,Adkins:2022omi,Ellis:2017kzh}, and so on. The first \mbox{QED} atom was discovered in 1951, which is the $e^+e^-$ bound state and named positronium~\cite{Deutsch:1951zza}; the second one was discovered in 1960, which is the $\mu^+e^-$ bound state and named muonium~\cite{Hughes:1960zz}. No other \mbox{QED} atom was found in the past 64 years.
New colliders are proposed to discover the true muonium~\cite{Fox:2021mdn,Gargiulo:2023tci} which decays into final states with electrons and photons~\cite{jentschura:1997bound,Brodsky:2009gx}.
The heaviest and smallest \mbox{QED} atom, a bound state of $\tau^+\tau^-$~\cite{Moffat:1975uw}, is named tauonium~\cite{Avilez:1977ai,Avilez:1978sa}, ditauonium~\cite{dEnterria:2022alo,Yu:2022hdt}, or true tauonium~\cite{dEnterria:2022ysg}. We classify the bound states of $\tau^+\tau^-$ following the charmonium spectroscopy~\cite{ParticleDataGroup:2022pth} simply as $J_\tau(nS)$, $\eta_\tau(nS)$, and $\chi_{\tau J}(nP)$ for the states with quantum numbers of $n^3S_1$, $n^1S_0$, and $(n+1)^{3}P_J$, respectively.
Here, $n$ is the radial quantum number, and $S$ and $P$ denote the orbital angular momentum of 0 and 1, respectively.
There were many theoretical studies in the literature since the discovery of the $\tau$ lepton. The spectroscopy of $\tau^+\tau^-$ atoms was studied in Ref.~\cite{dEnterria:2022alo}. The production of $\eta_\tau$ was considered in Refs.~\cite{dEnterria:2022ysg,Yu:2022hdt} and that of the process $e^+e^-\to J_\tau(1S)\to\mu^+\mu^-$ in Refs.~\cite{Avilez:1977ai,Malik:2008pn,dEnterria:2023yao}.
In addition, study of the $J_\tau(nS)$ via the processes $e^+e^-\to J_\tau(1S)\to {\rm light~hadrons}$~\cite{Avilez:1977ai,Avilez:1978sa} and $e^+e^-\to J_\tau(nS)\to \gamma\eta_\tau$ were proposed~\cite{Moffat:1975uw}.
Achievements have been made in the precise measurements of standard model and searches for exotic states at $e^+e^-$ colliders~\cite{BESIII:2014srs,Zhang:2018gol,Achasov:2019rdp,Anashin2007,Oset2023,He2022,Liu2023}.

\section{Methods and calculations}
In this Letter, we introduce a novel method for identifying $J_\tau(nS)$ by measuring the cross section ratio $\sigma(e^+e^-\to X^+ Y^-\slashed{E})/\sigma(e^+e^-\to\mu^+\mu^-)$. Here $X,\,Y=e$, $\mu$, $\pi$, $K$, or $\rho$, and $\slashed{E}$ is the missing energy due to unobserved neutrinos.
To suppress backgrounds with extra missing particles, one can require no additional photons in the event besides those from $\rho$ decays and the number of good charged tracks is exactly two for all the channels of $\tau\tau \to ee,~e\mu,~e\pi,~eK,~\mu\mu,~\mu\pi,~\mu K, \pi K,~\pi\pi,~KK,~e\rho,~\mu\rho,~{\rm and}~\pi\rho$. To remove background events from two-photon processes of $e^+e^- \to e^+e^-e^+e^-~{\rm and}~e^+e^-\to e^+e^-\mu^+\mu^-$, a lower limit of the transverse momentum of the charged tracks can be applied.
The process of $e^+e^-\to X^+ Y^-\slashed{E}$ can manifest in intermediate states in two distinct ways: $\tau^+ \tau^-$ open states above the double $\tau$ lepton mass threshold ($2m_\tau$) and $\tau^+ \tau^-$ bound states $J_\tau(nS)$ just below the threshold with a binding energy of $E_n \approx \alpha^2m_\tau/(4n^2) \approx 23.7/n^2$ keV.

By fitting the $\sigma(e^+e^-\to X^+ Y^-\slashed{E})/\sigma(e^+e^-\to \mu^+\mu^-)$ distribution in the vicinity of the $\tau$ pair production threshold with and without a $J_\tau(nS)$ atom component, we can identify its existence. We show below that the $\tau$ lepton mass measured with the process $e^+e^-\to\tau^+\tau^-\to X^+Y^-\slashed{E}$ will also be affected after considering the $J_\tau(nS)$ contribution.

The $\tau$ lepton mass is one of the fundamental parameters of the Standard Model. A precise $\tau$ mass measurement is essential to check the lepton flavor universality and constrain the $\nu_\tau$ mass~\cite{Marciano}. In the previous measurements of $e^+e^-\to \tau^+\tau^-$ cross sections at the \mbox{BESIII}~\cite{BESIII:2014srs,Zhang:2018gol,Achasov:2019rdp} and \mbox{KEDR} experiments~\cite{Anashin2007}, the contribution of $e^+e^-\to J_\tau(nS)$ was not included in theoretical calculation. This is not a problem at these experiments as the uncertainties are at 100~keV level (The current world average value of $\tau$ mass is $m^{\rm PDG}_\tau=(1776.86\pm 0.12)~{\rm MeV}$~\cite{ParticleDataGroup:2022pth} and the recent most precise measurement from \mbox{Belle II} experiment is $1777.09\pm 0.08\pm 0.11~{\rm MeV}$~\cite{Belle-II:2023izd}), but the effect should not be ignored in the next generation experiments which aim at a precision of one or two orders of magnitude better than previous ones.
In \mbox{BESIII} measurement~\cite{BESIII:2014srs}, the primary source of systematic uncertainty arises from the energy scale and efficiency. In \mbox{KEDR} measurement~\cite{Achasov:2019rdp}, the primary systematic uncertainty is attributed to efficiency and luminosity measurements. In the next generation of experiments, the statistical uncertainty is anticipated to decrease due to the large integrated luminosity, while the systematic uncertainty will be mitigated through the adoption of new fitting methods developed in this work and the application of advanced technologies to the detector and accelerator design.

The cross section $\sigma(e^+e^-\to X^+ Y^- \slashed{E})$ around the $\tau^+\tau^-$ production threshold is~\cite{BESIII:2014srs,Zhang:2018gol,Achasov:2019rdp,Anashin2007}
  \begin{eqnarray}
  \label{tautaucrosssecEX}
&&\sigma(W,m_{\tau},\Gamma_{\tau},\delta_W)=
\int_{m_{J_{\tau}}}^{\infty}dW^{\prime}
\frac{1}{\sqrt{2\pi}\delta_W}e^{-\frac{(W-W^{\prime})^2}{2\delta_W^2}} \nonumber \times\\
&&
\int_{0}^{1-\frac{m_{J_{\tau}}^2}{W^{\prime2}}}dx~ F(x,W^{\prime})
\frac{\bar{\sigma}(W^{\prime}\sqrt{1-x},m_{\tau},\Gamma_{\tau})}{|1-\Pi(W^{\prime}\sqrt{1-x})|^2}.
  \end{eqnarray}
Here, $W$ is the center-of-mass energy, $\delta_W$ is the center-of-mass energy spread, $F(x,W)$ is the initial state radiation factor~\cite{Kuraev:1985hb}, $\Pi(W)$ is the vacuum polarization factor~\cite{WorkingGrouponRadiativeCorrections:2010bjp}, and $\bar{\sigma}(W,m_{\tau},\Gamma_{\tau})$ is the Born cross section. With $J_\tau(nS)$ atoms included, \mbox{Eq.}~(\ref{tautaucrosssecEX}) differs from those given in Refs.~\cite{BESIII:2014srs,Zhang:2018gol,Achasov:2019rdp,Anashin2007}, where $2m_\tau$ is replaced by the ground state mass $m_{J_\tau}$ in the range of integration, the $\tau$ width $\Gamma_{\tau}$ is added as a variable, and the contribution of $J_\tau(nS)$ atoms ($\bar{\sigma}^{J_\tau}(W)$) is included in $\bar{\sigma}(W,m_{\tau},\Gamma_{\tau})$ as
  \begin{eqnarray}
\label{tautaucrosssec}
\bar{\sigma}(W,m_{\tau},\Gamma_{\tau})=\bar{\sigma}^{J_\tau}(W)+ \bar{\sigma}^{\rm con.}(W),
  \end{eqnarray}
where $\bar{\sigma}^{\rm con.}(W)$ is the cross section from the $e^+e^-\to \tau^+\tau^-$ continuum process and is calculated to the next-to-leading order (NLO) in the fine structure constant $\alpha$, as has been done for $\bar{\sigma}^{J_\tau}(W)$.

At NLO~\cite{Brodsky:2009gx,dEnterria:2022alo}, the cross section $\bar{\sigma}^{J_\tau}$ from narrow atoms is given by the Breit-Wigner function~\cite{ParticleDataGroup:2022pth}
\begin{eqnarray}\label{eq3}
\bar{\sigma}^{J_\tau}(W)
= \sum_n \frac{6 \pi^2  |1-\Pi(2m_\tau)|^2\left(1-\frac{3\alpha}{4\pi}\right)}{W^2  \Gamma^{J_\tau(nS)}_{\rm total} }\times \nonumber \\
\delta(W-m_{J_\tau(nS)})
\Gamma^{J_\tau(nS)}_{X^+ Y^-  \slashed{E}} \times \Gamma^{J_\tau(nS)}_{e^+e^-},
\end{eqnarray}
where $m_{J_\tau(nS)}=2m_\tau-E_n $ is the mass of $J_\tau(nS)$, ${|1-\Pi(2m_\tau)|^2}(1-3\alpha/4\pi)$ is recalled here since the initial state radiation factor and the vacuum polarization factor have been considered in \mbox{Eq.}~(\ref{tautaucrosssecEX}), $\Gamma^{J_\tau(nS)}_{X^+Y^-\slashed{E}}$ is the partial decay width of $J_\tau(nS)\to X^+ Y^-\slashed{E}$, and $\Gamma^{J_\tau(nS)}_{e^+e^-}$ is that of $J_\tau(nS)\to e^+e^-$. We have
  \begin{eqnarray}
 \Gamma^{J_\tau(nS)}_{e^+e^-}&=&\frac{ \alpha^5 m_\tau}{6 n^3 |1-\Pi(2m_\tau)|^2}\left(1-\frac{13\alpha}{4\pi}+C_{\rm coulomb}^{nS}\frac{\alpha}{\pi}\right),\nonumber \\
 \Gamma^{J_\tau(nS)}_{X^+ Y^- \slashed{E}}&=&2 \Gamma_\tau+\Gamma(J_\tau(nS)\to \gamma \chi_{\tau J}
 ),~{\rm and}   \nonumber\\
 \Gamma^{J_\tau(nS)}_{\rm total}&=&\Gamma^{J_\tau(nS)}_{ X^+ Y^-  \slashed{E}} +(2+R) \Gamma^{J_\tau(nS)}_{e^+e^-}.
  \end{eqnarray}

Futher detailed data are available in Table~\ref{tab:decayWidthofJtauN}, which is calculated to NLO in the fine structure constant $\alpha$~\cite{Dzikowski2019AGI,dEnterria:2022alo}.
Here, $2\Gamma_\tau$ is twice the
free-$\tau$ decay width (since the lifetime
of the bound state is about 10 times shorter than that of its components and relativistic corrections are neglected due to their minute contributions of the order of 10$^{-4}$). The uncertainties of $R$ and $\Gamma_\tau$ are included in the theoretical uncertainties.
The factor $(2+R)$ comes from $e^+e^-$, $\mu^+\mu^-$, and hadronic final states with $R_{\rm exp.}=2.342\pm 0.064$~\cite{BESIII:2021wib}, the total $\tau$ decay width $\Gamma_\tau=2.2674\pm 0.0039~{\rm meV}$~\cite{ParticleDataGroup:2022pth}, and $\Gamma(J_\tau(nS)\to \gamma \chi_{\tau J})$ is the $E1$ transition width (the annihilation decays of $\chi_{\tau J}$ are ignored since their contributions are smaller than $\bar{\sigma}^{J_\tau}(W)$ by a factor of $10^{-6}$).
With Green functions, the Coulomb corrections ($C_{\rm coulomb}^{nS}$) are calculated to be 5.804, 4.428, 3.810, 3.518, 3.358, 3.256, 3.186, 3.134, 3.093, and 3.061 for $n=1$, 2, 3, 4, 5, 6, 7, 8, 9, and 10, respectively.

Most of the NLO corrections of $\Gamma(J_\tau(nS)\to e^+e^-)$ come from the vacuum polarization factor $\Pi$. Then we get
\begin{eqnarray}
\bar{\sigma}^{J_\tau}(W)=(3.11\pm0.02)~\delta \left(\frac{W-2m_\tau+13.8~{\rm keV}}{1~{\rm MeV}}\right)~{\rm pb},
\end{eqnarray}
where $13.8~{\rm keV}=\sum_n E_n Br^{J_\tau(nS)}_{X^+ Y^-\slashed{E}} \Gamma^{J_\tau(nS)}_{e^+e^-}/\sum_n Br^{J_\tau(nS)}_{X^+ Y^- \slashed{E}} \Gamma^{J_\tau(nS)}_{e^+e^-}$ with $Br^{J_\tau(nS)}_{X^+Y^-\slashed{E}}$ being the branching fraction of $J_\tau(nS)\to X^+ Y^-\slashed{E}$. The uncertainty from $R$ is one order of magnitude greater than that from $m_\tau$ and $\Gamma_\tau$.

\begin{table}[htpb!]
\centering
\caption{\label{tab:decayWidthofJtauN}The decay data of $J_\tau(nS)$ in meV.
}
\resizebox{!}{!}{
\begin{tabular}{c ccccc}
\hline
 n    &$\Gamma^{J_\tau(nS)}_{e^+e^-}$& $2 \Gamma_\tau$
& $\Gamma^{J_\tau(nS)}_{E1} $ & $\Gamma^{J_\tau(nS)}_{\rm total}  $ &
$\Gamma^{J_\tau(nS)}_{e^+e^-} Br^{J_\tau(nS)}_{X^+Y^-\slashed{E}}$\\ \hline
 1    &6.484 & 4.535    & 0.0000 &32.695  &0.899\\
 2    &0.808 & 4.535    & 0.0000 & 8.044 &0.455\\
 3    &0.239 & 4.535    & 0.0072 & 5.573  &0.195 \\
\hline
$\sum_{n=1}^{\infty}$   &       &           &         &         &$1.795\pm0.012$ \\
\hline
\end{tabular}
}
\end{table}

\begin{figure}[htbp]
\includegraphics[width=\linewidth]{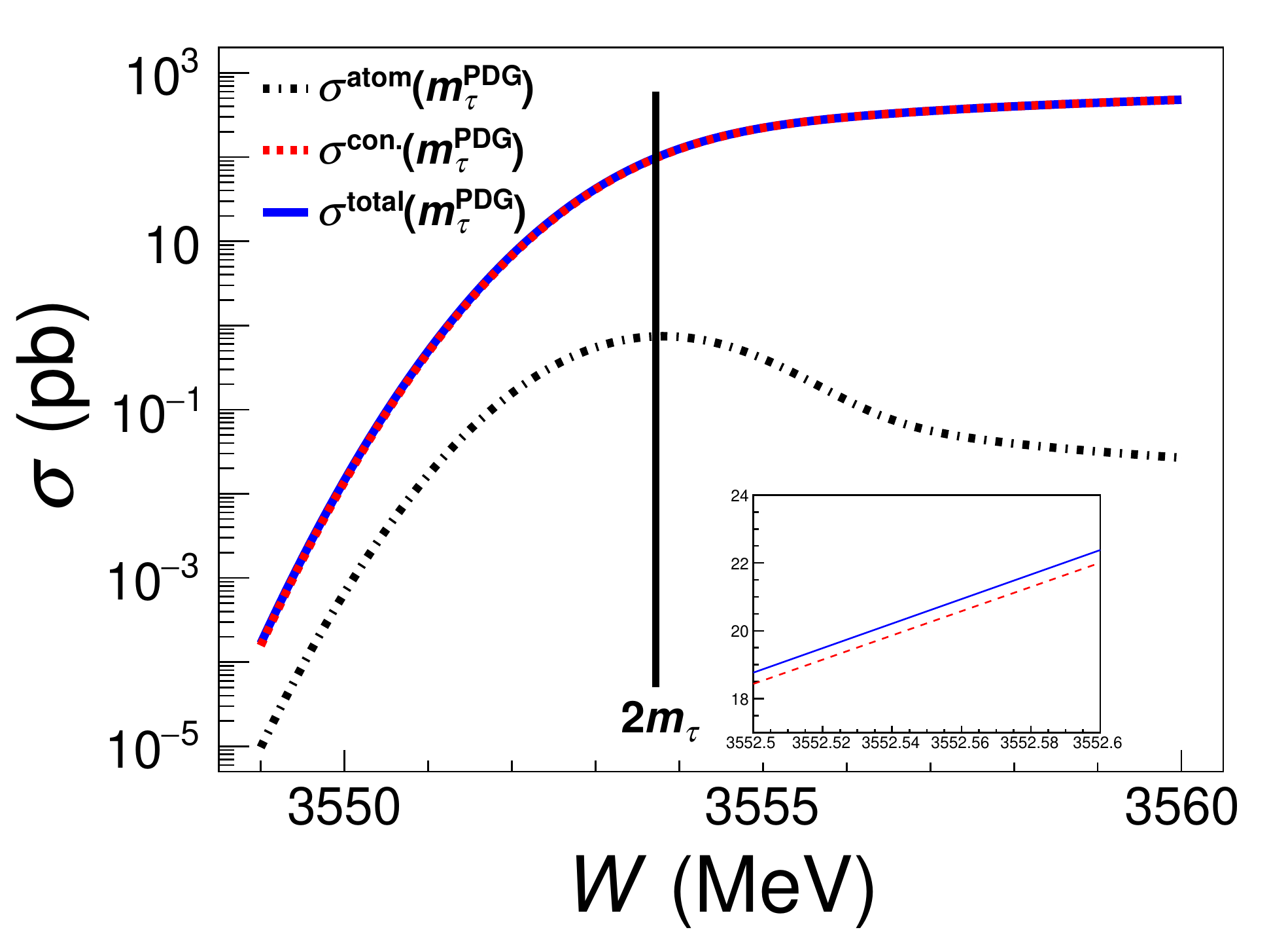}
\caption{\label{fig:crossCon.Res.Tot.} Cross sections $\sigma^{\rm atom}(m_\tau^{\rm PDG})$, $\sigma^{\rm con.}(m_\tau^{\rm PDG})$, and $\sigma^{\rm total}(m_\tau^{\rm PDG})$ as a function of center-of-mass energy $W$. The black vertical line shows the $\tau^+\tau^-$ mass threshold. The inset shows the $\sigma^{\rm con.}(m_\tau^{\rm PDG})$ and $\sigma^{\rm total}(m_\tau^{\rm PDG})$ as a function of $W$ in a small $W$ region of 3552.5 -- 3552.6 MeV.}
\end{figure}

We use the NLO cross sections $\bar{\sigma}^{\rm con.}(W)$ and take next-to-next-to-leading order (NNLO) corrections as uncertainties here~\cite{Voloshin:2002mv}. To reduce the uncertainties from the initial state radiation factor and the vacuum polarization factor in \mbox{Eq.}~(\ref{tautaucrosssecEX}), and that from the integrated luminosity~\cite{BESIII:2021wib}, we introduce $R_{X^+Y^-\slashed{E}}$, ratio of the cross sections, as
\begin{eqnarray}
\label{RnuXnuX}
R_{X^+Y^-\slashed{E}}(W,\delta_W,m_\tau)= \frac{\sigma(W,m_{\tau},\Gamma_{\tau},\delta_W)}{\sigma^{\mu^+\mu^-}(W,\delta_W)}.
  \end{eqnarray}
Here, $\sigma^{\mu^+\mu^-}(W,\delta_W)$ is calculated with $\bar{\sigma}^{\mu^+\mu^-}(W)=\frac{4\pi\alpha^2(1+3\alpha/4\pi )}{3 W^{2}}$ in \mbox{Eq.}~(\ref{tautaucrosssecEX}). The higher order correction terms, such as $9\alpha m_\mu^2/\pi W^2$ and $m_\mu^4/W^4$, are ignored because they are merely global factors of about $2\times10^{-5}$ here. With $m_\tau=m_\tau^{\rm PDG}$ and $\delta_W=1~{\rm MeV}$~\cite{Achasov:2023gey}, the cross sections $\sigma^{\rm atom}(m_\tau^{\rm PDG})$, $\sigma^{\rm con.}(m_\tau^{\rm PDG})$, and $\sigma^{\rm total}(m_\tau^{\rm PDG})$ are shown in \mbox{Fig.}~\ref{fig:crossCon.Res.Tot.}.

\section{Sensitivity of observing the $J_\tau(nS)$ and uncertainty of $\tau$ mass measuement}
Next, we estimate the sensitivity of observing the $J_\tau(nS)$ at a future high luminosity facility such as the super tau-charm facility (STCF) in China~\cite{Achasov:2023gey} and the super charm-tau factory (SCTF) in Russia~\cite{Anashin}. To determine which energy points are optimal for the study, we use the $\chi^2$ values per integrated luminosity as
\begin{eqnarray}
\label{Eq1}
\frac{\chi^2_i}{{\cal{L}}_i} = \frac{(\sigma^{\rm total}_i(m^{\rm PDG}_\tau) - \sigma_i^{\rm con.}(m_\tau))^2\cdot\varepsilon_{X^+ Y^-\slashed{E}}}{\sigma^{\rm total}_i(m^{\rm PDG}_\tau)},
\end{eqnarray}
where $\sigma^{\rm total}_i(m^{\rm PDG}_\tau)$ = $\sigma(W,m_{\tau},\Gamma_{\tau},\delta_W)$ in \mbox{Eq.}~(\ref{tautaucrosssecEX}) is the total cross section for energy point $i$ assuming $m_\tau=m^{\rm PDG}_\tau$, $\sigma_i^{\rm con.}$ is the cross section when only the continuum is included in \mbox{Eq.}~(\ref{tautaucrosssec}), $\varepsilon_{X^+ Y^-\slashed{E}}=8\%$ is the reconstruction efficiency of $e^+e^-\to X^+Y^-\slashed{E}$ events~\cite{BESIII:2014srs,Achasov:2023gey}, and ${\cal{L}}_i$ is the integrated luminosity. The reconstruction efficiency is estimated based on Monte Carlo simulations, where {\sc kkmc}~\cite{Jadach:1999vf,Jadach:2000ir} is used to simulate the production of $\tau$ pairs, and {\sc tauola}~\cite{Jadach:1990mz,Jadach:1993hs} is used to generate all the $\tau$ decay modes. Note that in the calculation of $\sigma_i^{\rm con.}(m_\tau)$, $m_{\tau}$ is allowed to vary so that $\frac{\chi^2_i}{{\cal{L}}_i}$ has different values at each energy point. Here, we choose the best solution by minimizing the value of $\Sigma_i\frac{\chi^2_i}{{\cal{L}}_i}$ within the region of $3.54<W<3.56$~GeV. In the end, we find the values of $\frac{\chi^2_i}{{\cal{L}}_i}$ are relatively large at $W=3552.56$ and $3555.83$~MeV, and that at $3552.56$~MeV is about half of that at $3555.83$~MeV. Besides the above two energy points, an additional energy point of $3549.00$~MeV is needed to obtain the whole lineshape of the $e^+e^-\to X^+Y^-\slashed{E}$ cross section.

We determine how large data samples are required in order to observe the $J_\tau(nS)$ at $W=3549.00$, $3552.56$, and $3555.83$~MeV by performing $10^5$ sets of simulated pseudoexperiments with the reconstruction efficiencies of $\varepsilon_{X^+ Y^-\slashed{E}}=(8.0\pm0.2)\%$~\cite{BESIII:2014srs, Achasov:2023gey} and $\varepsilon_{\mu^+\mu^-}=(45.00\pm 0.01)\%$~\cite{Ablikim:2002, Achasov:2023gey}, and other quantities used in \mbox{Eq.}~(\ref{eq3}).
Since the energy difference between energy points two and three is very small, we expect the efficiencies at these two points are very similar and almost 100\% correlated. The significance of the observation of the tauonium is independent of the uncertainty of the efficiency.
The numbers of expected events for $e^+e^-\to X^+ Y^-\slashed{E}$ and $e^+e^-\to\mu^+\mu^-$ in simulated data samples are determined by $N^{\rm data}_{X^+ Y^-\slashed{E}} = \sigma^{\rm total}(m^{\rm PDG}_\tau)\cdot{{\cal{L}}} \cdot \varepsilon_{X^+ Y^-\slashed{E}}$ and $N^{\rm data}_{\mu^+\mu^-} = \sigma^{\mu^+\mu^-}\cdot{{\cal{L}}} \cdot \varepsilon_{\mu^+\mu^-}$. The statistical uncertainties of $N^{\rm data}_{X^+ Y^-\slashed{E}}$ and $N^{\rm data}_{\mu^+\mu^-}$ are the square roots of them.
For $e^+e^-\to X^+ Y^-\slashed{E}$ at $W=3549.00$~MeV, since the signal yield ($N$) is small, the statistical uncertainty of $N$ is estimated with the Bayesian approach implemented in the Bayesian Poisson Upper Limit Estimator at a 68.27\% confidence level~\cite{Zhu:2007zza}, where the number of expected background events is zero. The numbers of expected events and the statistical uncertainties for $e^+e^-\to X^+ Y^-\slashed{E}$ and $e^+e^-\to\mu^+\mu^-$ in the simulated data samples are summarized in Table~\ref{tab:eventsOFmumuANDxxe}, where the integrated luminosities are optimized and determined based on the $\chi^2$ value to estimate the $J_\tau(nS)$ signal significance reaching a 5$\sigma$ level (discussed below). For each set of pseudoexperiment, we generate randomly the numbers of events ($N^{\rm data}_{X^+ Y^-\slashed{E},\,i}$ and $N^{\rm data}_{\mu^+\mu^-,\,i}$, $i=1$, $2$, and $3$) according to Poisson distributions.

\begin{table}[htbp]
\centering
\footnotesize
\caption{\label{tab:eventsOFmumuANDxxe} Numbers of $e^+e^-\rightarrow X^+Y^-\slashed{E}$ and $\mu^+\mu^-$ events and their statistical uncertainties in the pseudoexperiments with $m_\tau=m_\tau^{\rm PDG}$.}
\resizebox{!}{!}{
\begin{tabular}{ccccc} \hline
$i$  &${\cal L}_i$ $({\rm fb}^{-1})$&  $W_i$ (MeV) & $N^{\rm data}_{X^+Y^-\slashed{E},~i}$   & $ N^{\rm data}_{\mu^+\mu^-,~i}$            \\ \hline
1    &    5                   &  3549.00       & $0.1^{+1.2}_{-0.1}$                  & $(1.1764\pm0.0003)\times 10^7$    \\
2    &    500                 &  3552.56       & $(8.772\pm0.009)\times 10^5$         &  $(1.17394\pm0.00003)\times 10^9$     \\
3    &    1000                &  3555.83       & $(2.4052\pm0.0005)\times 10^7$~~ &  $(2.34331\pm0.00005)\times 10^9$ \\\hline
\end{tabular}
}
\end{table}

\begin{figure}[htbp]
\includegraphics[width=\linewidth]{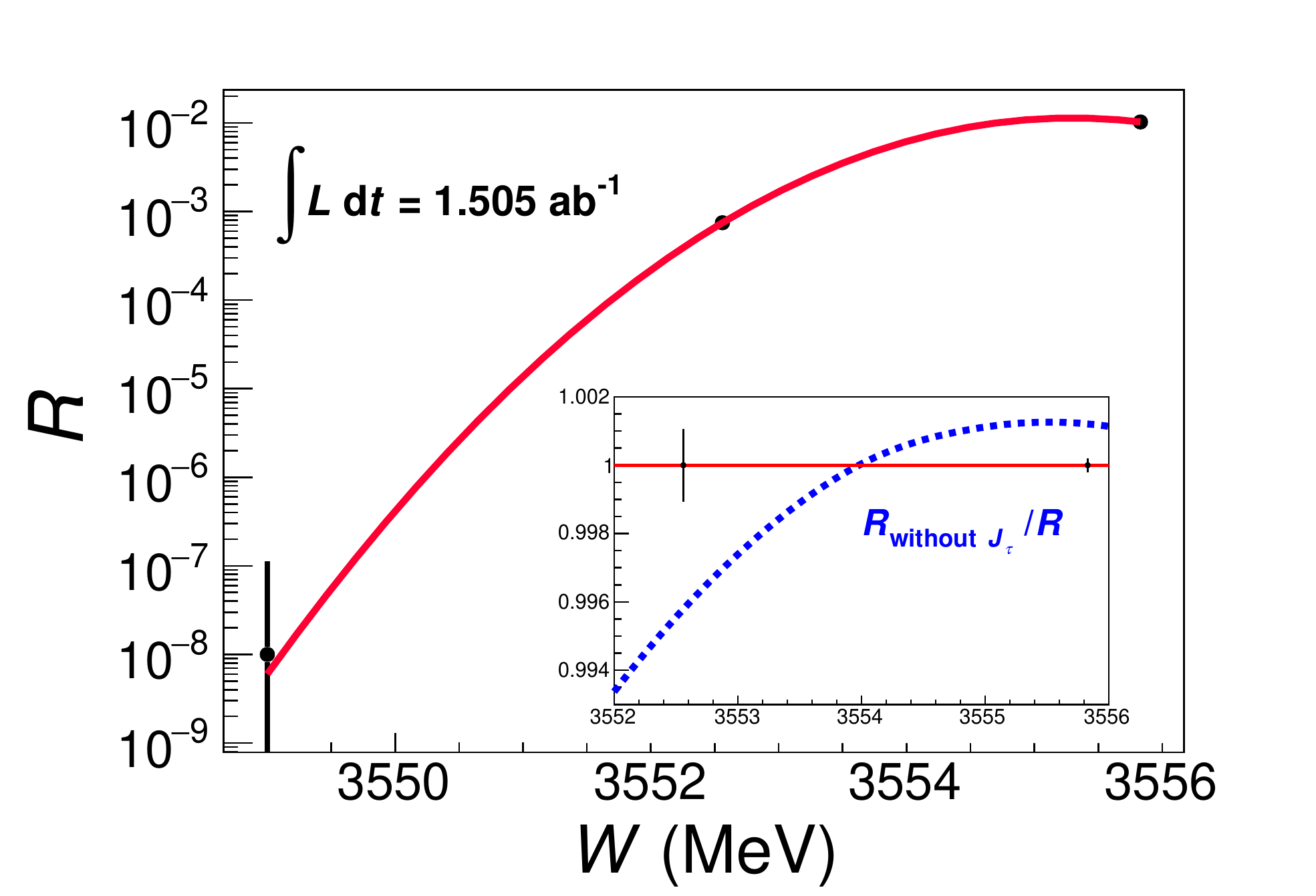}
\put(-240, 150){\large \bf (a)}

\includegraphics[width=\linewidth]{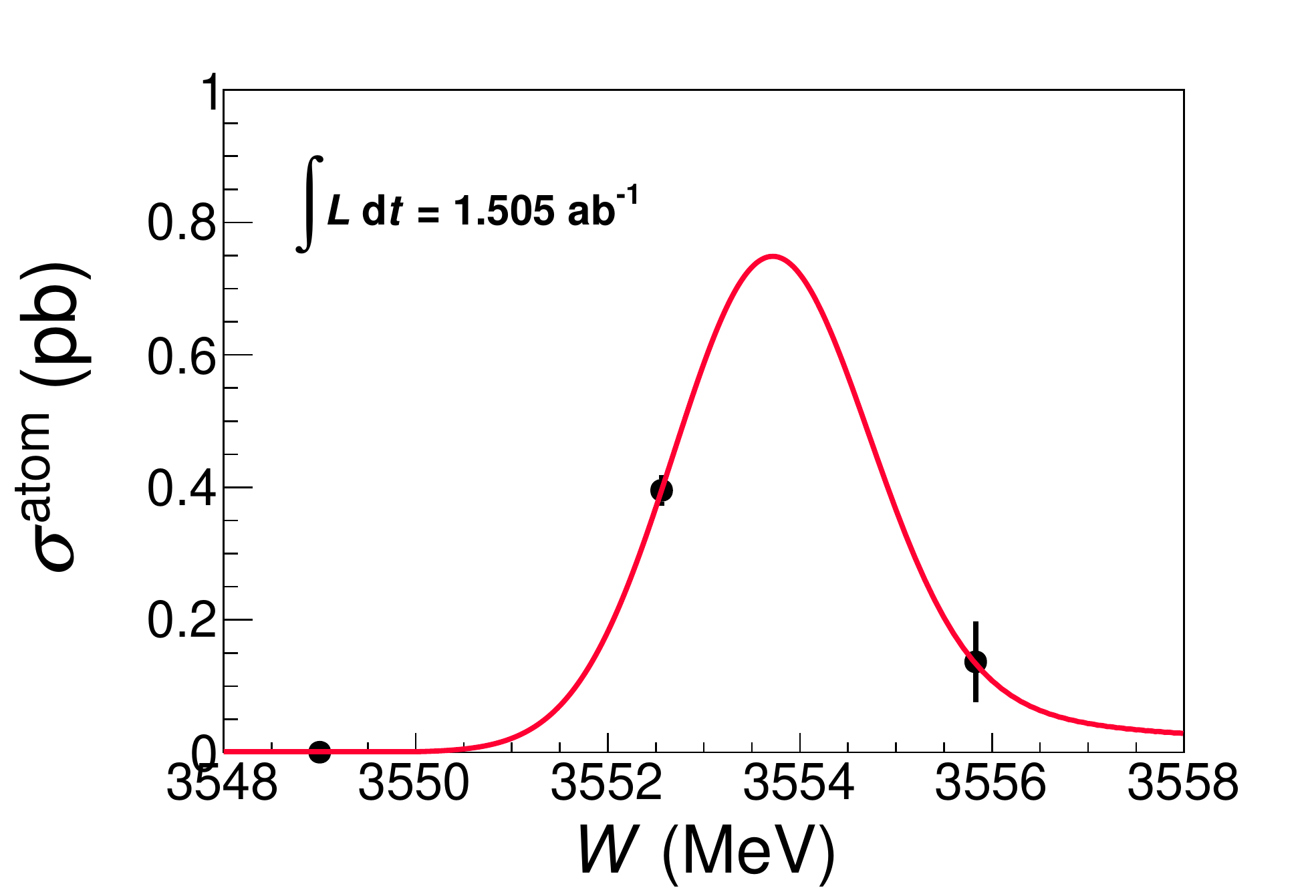}
\put(-240, 150){\large \bf (b)}
\caption{\label{fig2} (a) The fit to ${\cal R}^{\rm data}$ from one set of pseudoexperiment data, and (b) the  $\sigma^{\rm atom}$ contribution from the fit. The dots with error bars are the pseudoexperiment data, and the red curves display the best fit. The blue dashed line in the inset of (a) shows ${\cal R}_{\rm without~J_\tau}/{\cal R}$.}
\end{figure}

A least-square fit is applied to each set of the pseudoexperiments with
  \begin{eqnarray}
\label{Eq2}
\chi^2 = \sum_{i=1}^{3}\left(\frac{{\cal R}^{\rm data}_i-{\cal \hat{R}}_i(m_\tau)}
{\Delta {\cal R}^{\rm data}_i} \right)^2,
\end{eqnarray}
where ${\cal R}^{\rm data}_i = \frac{N^{\rm data}_{X^+ Y^-\slashed{E},\,i}}{N^{\rm data}_{\mu^+\mu^-,\,i}}$ and $\Delta {\cal R}^{\rm data}_i$ is its statistical uncertainty calculated from those of $N^{\rm data}_{X^+ Y^-\slashed{E},\,i}$ and $N^{\rm data}_{\mu^+\mu^-,\,i}$; ${\cal \hat{R}}_i(m_\tau)$ is the expected ratio at the $\tau$ mass $m_\tau$ to be determined from the fit. The fit to one pseudoexperiment is shown in \mbox{Fig.}~\ref{fig2}a, and the corresponding contribution from the $J_\tau(nS)$ atom cross section ($\sigma^{\rm atom}$) is shown in \mbox{Fig.}~\ref{fig2}b. For $10^5$ sets of simulated pseudoexperiments, the average value of $\chi^2/{\rm ndf}$ is 0.7/2, where ndf is the number of degrees of freedom. This indicates a very good fit to the simulated data samples.

\begin{figure}[htbp]
\includegraphics[width=\linewidth]{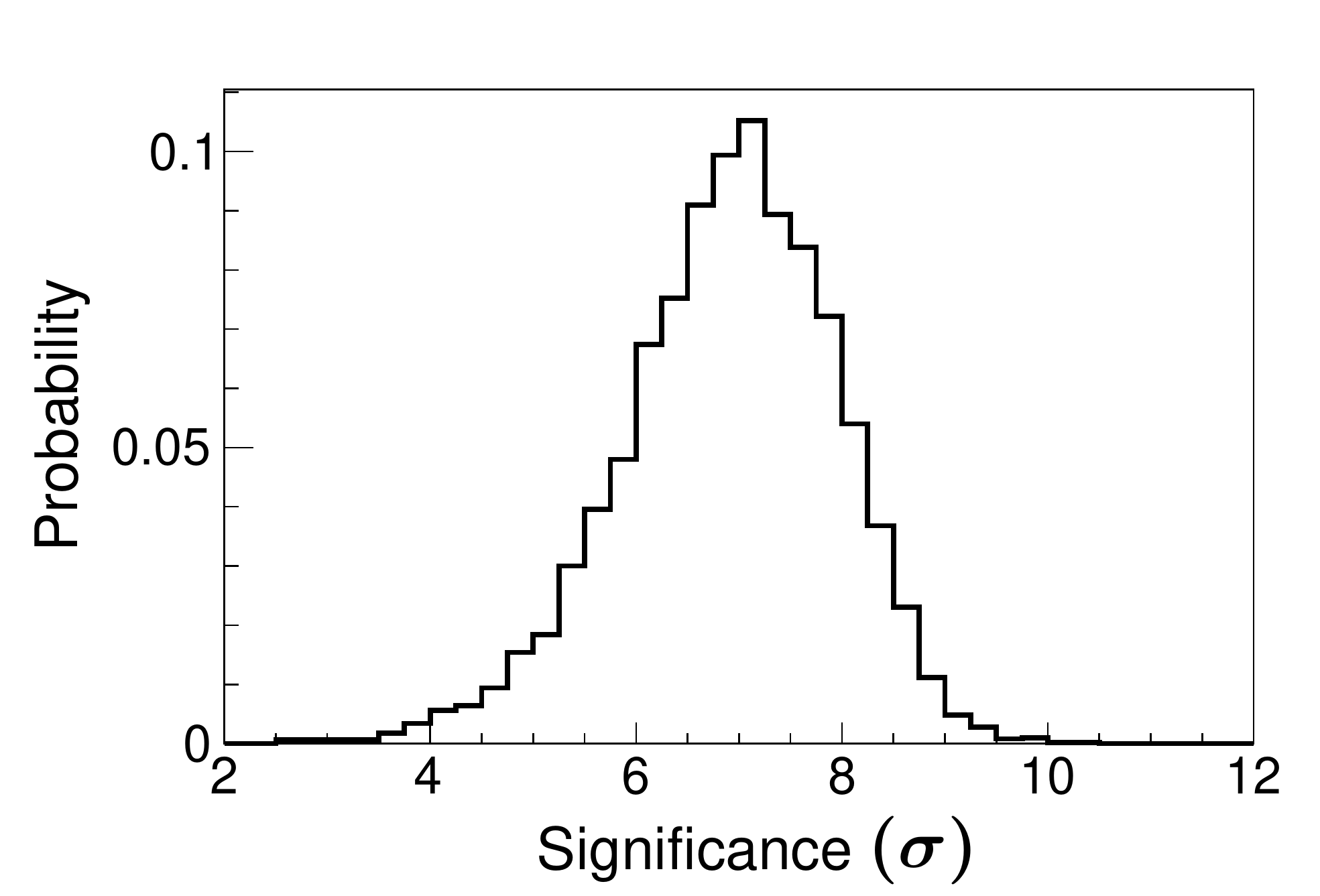}
\caption{\label{significance} Normalized distribution of the statistical significance of the $J_\tau(nS)$ signals in all the pseudoexperiments.}
\end{figure}

By removing the $J_\tau(nS)$ atom contribution in calculating ${\cal \hat{R}}_i(m_\tau)$ and refiting the data, we find a much poorer fit quality (the average value of $\chi^2/{\rm ndf}$ is 51/2 for the $10^5$ sets of simulated pseudoexperiments) and the difference in the $\chi^2$s measures the statistical significance of the $J_\tau(nS)$ signals. Figure~\ref{significance} shows the normalized distribution of the statistical significances in all the pseudoexperiments. We conclude that in the scenario of taking 5~fb$^{-1}$ data at 3549.00~MeV, 500~fb$^{-1}$ at 3552.56~MeV, and 1000~fb$^{-1}$ at 3555.83~MeV as indicated in Table~\ref{tab:eventsOFmumuANDxxe}, we have a 96\% chance of discovering the $J_\tau(nS)$ with a statistical significance larger than $5\sigma$ and an almost 100\% chance of observing it with a significance larger than $3\sigma$. These data samples correspond to about 350 and 175 days' data taking time at the STCF~\cite{Achasov:2023gey} and SCTF~\cite{Anashin}, with designed instantaneous luminosities of $0.5\times 10^{35}$ and $1.0\times10^{35}$~cm$^{-2}$s$^{-1}$, respectively. Here, we assume the efficiency and $\delta_W$ at the SCTF are the same as those at the STCF. If the $\delta_W$ is reduced to $0.1~{\rm MeV}$, the required integrated luminosity of the data is only $66~{\rm fb}^{-1}$.
In order to minimize the impact of beam energy instability and detector performance, it is advisable to frequently measure the beam energy and collect data at two high energy points in multiple rounds, with the integrated luminosity ratio of 1:2, rather than completing data collection at one energy before moving to the other.

The background cross section can be measured at $W_1$. The background cross section $\sigma_B=0^{+0.12}~{\rm pb}$ is obtained at \mbox{BESIII}~\cite{BESIII:2014srs}. If we set $\sigma_B=0.12/2~{\rm pb}$ in data, and $\sigma_B$ and $m_\tau$ float in the fit, $\Delta\chi^2_{J_\tau}/{\rm ndf}=45.4/1$ with the significance of 6.7$\sigma$ is obtained, where $\Delta \chi_{J_\tau}^{2}=\chi_{{\rm without}~J_\tau}^{2}-\chi_{J_\tau}^{2}$.
Taking into account the non-$\tau$ background, i.e. changing the number of background events to be $10_{-10}^{+120}$ (corresponding to an integrated luminosity of $500~{\rm fb^{-1}}$) for the second point and $20_{-20}^{+240}$ (corresponding to an integrated luminosity of $1000~{\rm fb^{-1}}$)  for third energy point, an average signal significance of the $J_{\tau}$ is reduced by $0.02\sigma$ from $6.8\sigma$. The ratios of the numbers of non-$\tau$ backgrounds including the uncertainty based on the \mbox{BESIII} cross section determination~\cite{BESIII:2014srs} relative to the numbers of $e^+e^-\rightarrow X^+Y^-\slashed{E}$ are $1.5\times10^{-4}$ and $1.1\times10^{-5}$ at $W$ = 3552.56 and 3555.83 MeV, respectively. Therefore, the non-$\tau$ backgrounds are negligible.

With these data samples, we obtain a high precision $\tau$ mass measurement. The above fit yields
$$
m_{\tau} = (1~776~860.00\pm0.25~({\rm stat.})\pm0.99~({\rm syst.}))~{\rm keV},
$$where the first and second uncertainties are statistical and systematic, respectively.
The fit with the $J_\tau(nS)$ contribution removed gives  a shift of $-4$~keV relative to the nominal fit with both the bound state and continuum contributions. This shift is about a factor of 4 larger than the total uncertainty and should not be ignored in the future high precision measurements.

\section{Systematic uncertainties}
The systematic uncertainties in the $\tau$ mass measurement are listed in Table~\ref{sys}.
The uncertainty of the center-of-mass energy scale $W$ is estimated according to that achieved twenty years ago at the VEPP-4M, which had a characteristic uncertainty of 1.5 keV in the beam energy in the $\psi(2S)$ and $J/\psi$ mass scan experiments using the resonant depolarization method~\cite{KEDR:2003bik,Anashin:2015rca}.
The uncertainty of $W_2~(W_3)$ is estimated to be $1.5\sqrt{2}=2.12$~keV, leading to an uncertainty of  $0.72~(0.35)$~keV in  $m_\tau$.

The energy spread can be measured from experimental data directly.
In the previous measurements, the $m_\tau$ uncertainties from energy spread and energy scale are 16 keV and $^{+22}_{-86}$ keV from \mbox{BESIII}~\cite{BESIII:2014srs}, and 25 keV and 40 keV from \mbox{KEDR}~\cite{Anashin2007}.
Therefore, we conservatively take the maximum ratio of $16/22\sim 0.73$, and find the uncertainty in the $m_\tau$ measurement from energy spread
is $0.73\times\sqrt{0.72^2+0.35^2}$ = 0.59 keV, where the 0.72 and 0.35 keV are the $m_\tau$ uncertainties from energy scales at $W_2$ and $W_3$, respectively. In the $m_\tau$ measurements at the \mbox{BESIII}~\cite{BESIII:2014srs} and the \mbox{KEDR}~\cite{Anashin2007}  experiments, $\delta_W$ is fit from $\delta_{m _{\psi(2S)}}$ and $\delta_{m_{J/\psi}}$, where $\delta_{m _{\psi(2S)}}$ and $\delta_{m_{J/\psi}}$ are free parameters and are obtained from fits to the measured $\psi(2S)$ and $J/\psi$ excitation curves~\cite{Anashin:2015rca}. If we consider $\delta_W$ as a free parameter in the $m_\tau$ fit function in \mbox{Eq.}~(\ref{Eq2}), the $0.27$~keV uncertainty of $m_\tau$ is given through $\Delta m_\tau^2|_{\chi^2=2}-\Delta m_\tau^2|_{\chi^2=1}$.

Considering the improvement of the particle identifications for $e$, $\mu$, $\pi$, $K$, and $\rho$ candidates at the future experiments, we expect the efficiency and its uncertainty of $\varepsilon_{X^+ Y^-\slashed{E}}=(8.0\pm 0.2)\%$.
We change the efficiency by $\pm1\sigma$, and find a change of 0.04 keV on the $\tau$ mass compared with the nominal result.

By replacing the NLO correction with the NNLO correction in the calculation of the $e^+e^-\to X^+Y^-\slashed{E}$ cross sections, we find the $\tau$ mass changes by 0.07~keV which is included as the uncertainty due to the theoretical accuracy.
Since we perform the fit to the ratio of observed $e^+e^-\to X^+Y^-\slashed{E}$ and $\mu^+\mu^-$ events, the uncertainty from the integrated luminosity is cancelled.
For the first energy point with a small expected number of events and large statistical uncertainty, we enlarge the uncertainty on the $N^{\rm data}_{X^+Y^-\slashed{E}}$ by a factor of three, and the resulting $\tau$ mass does not change.

Assuming all these sources are independent, we add them in quadrature to obtain the total systematic uncertainty, which is listed in Table~\ref{sys}.
Taking into account the systematic uncertainties mentioned above, we obtain an average signal significance of the $J_\tau$ of $6.7\sigma$, which will be $6.8\sigma$ if the systematic uncertainties are not taken into account.

\begin{table}[htbp]
\centering
\caption{\label{sys}
 The systematic uncertainties of the $m_{\tau}$ ($\sigma_{m_{\tau}}$) in keV.
}
\resizebox{!}{!}{
\begin{tabular}{c c}
\hline
Sources                 &  $\sigma_{m_{\tau}}$ (keV)   \\ \hline
Energy scale of $W_2$     &  0.72   \\
Energy scale of $W_3$     &  0.35   \\
Energy spread $\delta_W$        &  0.59   \\
Efficiency     &  0.04   \\
Theory     &  $0.07$   \\
 \hline
Systematic uncertainties       &  0.99   \\
\hline
\end{tabular}
}
\end{table}

\begin{figure}[htbp]
  \includegraphics[width=\linewidth]{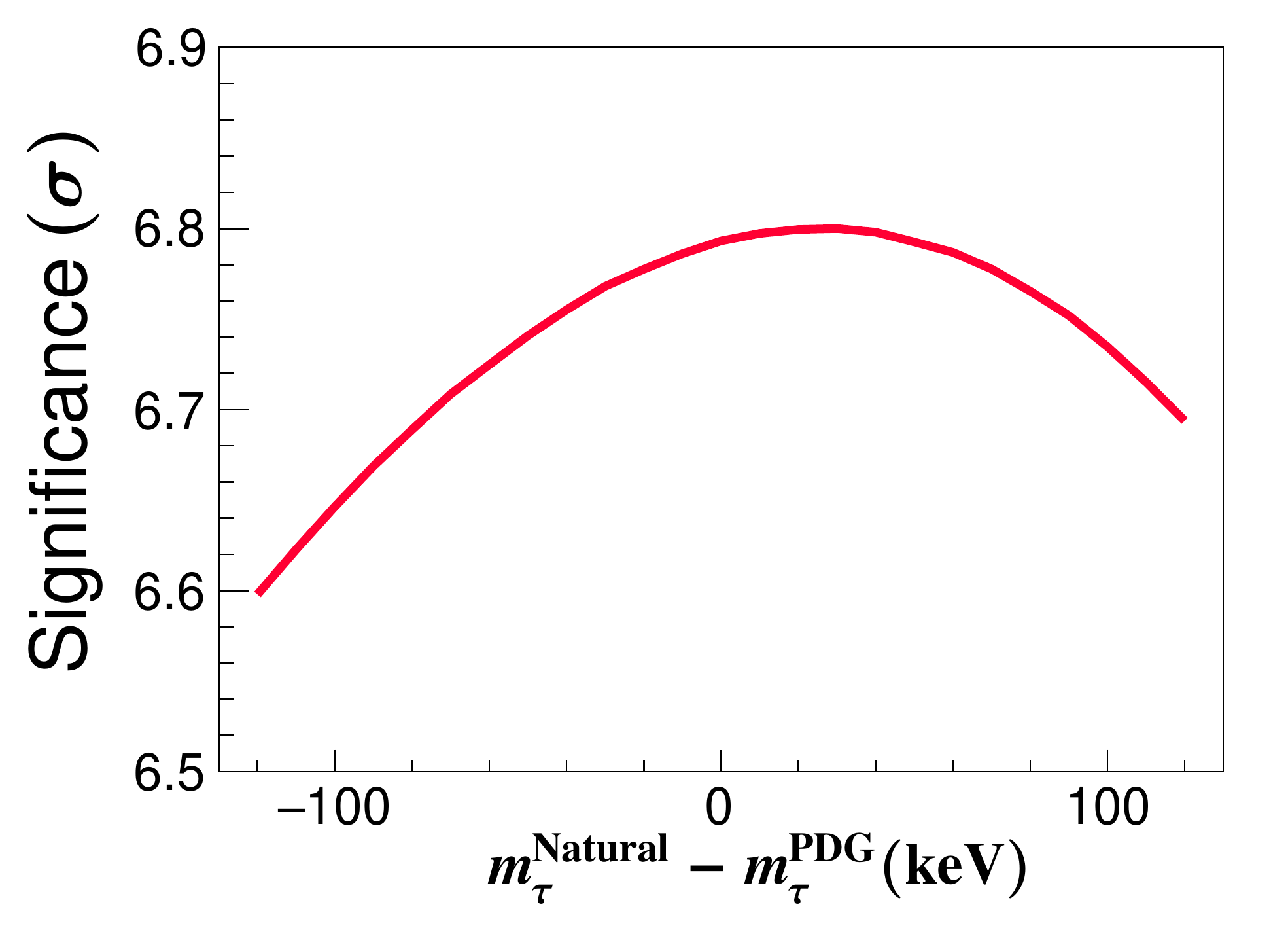}
  \caption{\label{Fig:NaturalTauMassSignificance} The significance of $J_\tau(nS)$ 
  as a function of $m_\tau^{\rm Natural} -m_\tau^{\rm PDG}$.}
\end{figure}

The value of $m_\tau^{\rm Natural}$ may be different from the central value of $m^{\rm PDG}_\tau=1776.86~{\rm MeV}$
~\cite{ParticleDataGroup:2022pth}. We examine the  statistical significance of the $J_\tau(nS)$ signals  as a function of the difference $m_\tau^{\rm Natural} -m_\tau^{\rm PDG}$ as shown in \mbox{Fig.}~\ref{Fig:NaturalTauMassSignificance}, and find that the uncertainty in $m_\tau$ remains unchanged, since it is mainly determined by the beam properties.

\section{Conclusion}
To conclude, the novel method proposed in this Letter is summarized as follows:
(1) In contrast to the process $e^+e^- \to J_\tau \to \mu^+ \mu^-$ proposed in Ref.~\cite{dEnterria:2023yao}, the continuum contributions are much smaller and the selected $\tau$ pair candidate sample is very pure in the process $e^+e^- \to J_\tau \to \tau^+ \tau^-$. The signal to background ratio in $e^+e^- \to J_\tau \to \tau^+ \tau^-$ is improved drastically.
(2) We propose to measure the relative rate ${\cal R} = \frac{N_{X^+ Y^-\slashed{E}}}{N_{\mu^+\mu^-}}$ rather than the absolute cross section so that the uncertainties are controlled at a low level since those in VP, ISR, and luminosity determinations are canceled.
(3) $m_\tau$ is taken as a free parameter to be extracted from the experimental data. A high precision $m_\tau$ measurement can be achieved at the same time.

In summary, we show that the $\tau^+\tau^-$ atom with $J^{PC}=1^{--}$, $J_\tau$, can be observed with a significance larger than 5$\sigma$ with a $1.5~{\rm ab^{-1}}$ data sample at the proposed high luminosity experiments STCF and SCTF, by measuring the cross section ratio of the processes $e^+e^-\to X^+ Y^-\slashed{E}$ and $e^+e^-\to \mu^+\mu^-$. With the same data sample, the $\tau$ lepton mass can be measured with a precision of 1~keV, a factor of about 100 improvement over the existing world best measurements.

\section*{Acknowledgments}
We thank Profs. Kuang-Ta Chao and Hua-Sheng Shao for valuable and helpful discussions. This work is supported in part by National Key Research and Development Program of China under Contract No.~2020YFA0406300, National Natural Science Foundation of China (NSFC) under contract No.~11975076, No.~12161141008, No.~12135005, No.~12075018, No.~12005040, and No.~12335004; and the Fundamental Research Funds for the Central Universities Grant No.~RF1028623046.

\section*{Author contributions}

Yu-Jie~Zhang proposed this project. Jing-Hang~Fu and Yu-Jie~Zhang did the calculations. Sen~Jia, Xing-Yu~Zhou, Cheng-Ping~Shen, and Chang-Zheng~Yuan carried out the experiments. Cheng-Ping~Shen and Chang-Zheng~Yuan checked the results and resolved physical and technical issues. 
Yu-Jie~Zhang, Cheng-Ping~Shen and Chang-Zheng~Yuan supervised throughout this work.
All authors contributed to the writing of the manuscript.

\end{document}